\documentclass[
aps,nofootinbib,showpacs,showkeys,preprint
,tightenlines ] {revtex4}

\usepackage{epsf,epsfig,subfigure,graphicx,amsmath,amssymb}
\usepackage{color}

\newcommand{\dis}[1]{\begin{equation}\begin{split}#1\end{split}\end{equation}}
\newcommand{\be}{\begin{equation}}
\newcommand{\ee}{\end{equation}}

\newcommand{\eq}[1]{Eq.~(\ref{#1})}

\newcommand{\tev}{\,\textrm{TeV}}

\def\bea{\begin{eqnarray}}
\def\eea{\end{eqnarray}}

\begin{document}

\title{\Large\bf U(1)$_{B_1+B_2-2L_1}$ mediation for the natural SUSY \\and the anomalous muon $g-2$
}

\author{Ji-Haeng Huh$^{1}$\footnote{email: jhhuh@physics.ucla.edu} 
and Bumseok Kyae$^{2}$\footnote{email: bkyae@pusan.ac.kr}
}
\affiliation
{$^{1}$Department of Physics, University of California, Los Angeles, CA 90095, USA \\ 
$^{2}$Department of Physics,  Pusan National University, Busan 609-735, Korea}

\begin{abstract}

We propose a U(1)$^\prime$ mediated supersymmetry (SUSY) breaking, in which U(1)$^\prime$ is identified with U(1)$_{B_1+B_2-2L_1}$. The U(1)$_{B_1+B_2-2L_1}$ gauge symmetry, which is anomaly-free with the field
contents of the minimal supersymmetric standard model, assigns $\pm 1/3$ charges to
the first and second generations of the quarks, and $\mp 2$ to the first generation of the leptons.
As a result, the first two generations of squarks acquire masses of about 7 TeV,
and the first generation of the sleptons do those of 40 TeV, respectively, in the presence of one or three pairs of extra vector-like matter $\{{\bf 5},\overline{\bf 5}\}$.
Non-observation on extra colored particles below 1 TeV at the large hadron collider, and also the flavor
violations such as $\mu^-\rightarrow e^-\gamma$ are explained.
By virtue of such a gauge symmetry, proton stability can be protected.
The other squarks and sleptons as well as the gauginos can obtain masses of order $10^{2-3}$ GeV
through the conventional gravity or gauge mediated SUSY breaking mechanism.
The relatively light smuon/sneutrino and the neutralino/chargino could be responsible for the $(g-2)_\mu$
deviated from the standard model prediction.
The stop mass of $\sim 500~{\rm GeV}$ relieves the fine-tuning problem in the Higgs sector.
Two-loop effects by the relatively heavy sfermions can protect the smallness of the stop mass from the radiative correction by
the heavy gluino ($\gtrsim 1~{\rm TeV}$).
Extra vector-like matter can enhance the radiative corrections to the Higgs mass up to 126 GeV, and induce
the desired mixing among the chiral fermions after U(1)$_{B_1+B_2-2L_1}$ breaking.

\end{abstract}

\pacs{12.60.Jv, 14.80.Ly, 11.25.Wx, 11.25.Mj}

\keywords{U(1)$^\prime$ mediation, effective SUSY, muon anomalous magnetic moment}

\maketitle


\section{Introduction}

Recently, ATLAS and CMS have announced the discovery of the standard model (SM) Higgs(-like) boson in the
$125-126 ~{\rm GeV}$ invariant mass range \cite{ATLAS,CMS}.  Thus, the SM seems to have been almost completely confirmed as the basic theory describing the nature.
So far, any evidence beyond the SM
including supersymmetry (SUSY) has not appeared yet at the large hadron collider (LHC).
It implies that the theoretical puzzles raised in the SM such as the gauge hierarchy problem still remain unsolved \cite{text}.

In fact, 126 GeV is too large for the mass of Higgs boson in the minimal SUSY SM (MSSM), 
if we conservatively keep the original motivation of introducing SUSY at the EW scale. In the MSSM
the observed 126 GeV Higgs mass requires a too heavy stop mass ($\widetilde{m}_t\gtrsim$ a few TeV) for
enhancing the radiative Higgs mass \cite{twoloop},
which compels the soft parameters to be finely tuned to match the $Z$ boson mass at the minimum of the
Higgs potential \cite{text}.
Thus, an excessively heavy stop mass spoils the status of SUSY as a solution of
the fine-tuning problem associated with the SM Higgs boson mass.

For naturalness of the Higgs boson mass, thus,  we need a relatively light stop.
At the moment, fortunately the stop mass bound is much lower compared to those of the other squarks,
$\widetilde{m}_t\gtrsim 500$ GeV, which provides just
$\Delta m_h^2|_{\rm top}\gtrsim (71~{\rm GeV})^2$  through one-loop radiative correction together with the top quark.
As a result, we need more ingredients beyond the MSSM for raising the Higgs mass except the stop:
for explaining the 126 GeV Higgs mass, $\Delta m_h^2|_{\rm new}\sim (89-51~{\rm GeV})^2$
for ${\rm tan}\beta=2-50$ should be supplemented by extending the MSSM,
if the stop mass is taken to be just around 500 GeV in order to minimize the fine-tuning in the Higgs sector, avoiding the experimental bound on it\footnote{For heavy LSP ($\gtrsim 350~{\rm GeV}$), $\widetilde{m}_t$ is not constrained yet \cite{stopmass}.} (see, for instance, Refs.~\cite{KP,KS2,KS1}).

The fact that any SUSY particles have not been observed yet at the LHC means that the masses of new colored particles absent in the SM would be quite heavier than 1 TeV.
Here we should note that such a bound is still applied only to the first and second generations of squarks and to the gluino: the constraint on the third generation is just around 500 GeV, as mentioned above.
Fortunately, such heavy squark masses do not affect much the Higgs mass
because of their small Yukawa couplings. Moreover, heavy masses of the first and second
generations of squarks are helpful for avoiding the flavor problem expected in
the gravity mediation scenario.
If we insist on the stop mass being $\sim$ 500 GeV at the electroweak\,(EW) scale, however,
gluino heavier than 1 TeV at the EW scale would drive the squared mass of the stop negative
at a higher energy scale through the renormalization group (RG) effects.
It means that the stop becomes much heavier than 500 GeV at low energies, if we take it positive at high energy scales.
To avoid color breaking at a higher energy scale, and keep the small stop mass at the EW scale, we need to
somehow compensate the heavy gluino effect in the RG equation.

Brookhaven National Laboratory (BNL) reported a remarkable result on $(g-2)$ of the muon ($\mu$) \cite{BNL}
which is deviated from the SM prediction  \cite{SMg-2,hadUncertain} by $3.3\sigma-3.6\sigma$,
\dis{
\Delta (g-2)_\mu =(g-2)_\mu^{\rm exp}-(g-2)_\mu^{\rm SM}=(26.1\pm 8.0)\times 10^{-10} .
}
Even though the statistical significance is not strong enough in addition to the theoretical uncertainty on the hadronic effects, still this discrepancy may hint new physics beyond the SM.
If it originates from SUSY, relatively light smuon/sneutrino and neutralino/chargino are needed to give this order of $(g-2)_\mu$ \cite{MSSMg-2,Rviolg-2}. 
In contrast to the squarks, the mass bounds on sleptons at the LHC are
less severe \cite{PDG}. As pointed out in Ref.~\cite{KS2}, moreover, the extra vector-like leptons lighter
than $\sim 500~{\rm GeV}$
is very helpful for explaining 126 GeV Higgs mass with $\widetilde{m}_t\sim 500~{\rm GeV}$
but without assuming a large mixing between the left- and the right-handed stops.
Their order-one Yukawa coupling to the MSSM Higgs boson can enhance the radiative Higgs mass.
The Landau-pole problem associated with the order-one Yukawa coupling can be avoided
by introducing an extra gauge symmetry, under which {\it only the new vector-like leptons are charged}.

In view of the recent experimental data at the LHC, the ``effective SUSY'' (or ``more minimal SUSY'')
\cite{effSUSY,RecentEffsusy} and ``split SUSY'' \cite{splitSUSY} scenarios look promising.
According to the effective SUSY, the first two generations of superpartners are required to be about
$5-20\, \tev$ in order to avoid the SUSY flavor and SUSY CP problems, while the third ones
and gauginos can be lighter than 1 TeV. However, the masses heavier than 22 TeV for the first two
generations drive the stop mass squared negative at the EW scale through the two-loop RG effects,
if the third ones are lighter than 4 TeV, and if such soft parameters are generated around the grand unification (GUT) scale ($\approx 2\times 10^{16}~{\rm GeV}$) \cite{tachyon}.
We note here that such a two-loop RG effect by heavy superpartners on lighter superpartners can
be utilized to keep the stop mass squared positive at higher energy scales, {\it when the gluino is also heavy enough} ($\gtrsim 1~{\rm TeV}$).
Namely, it can compensate the heavy gluino effect on the stop mass we mentioned above, keeping the light stop mass at low energy. 
In this Letter, we will discuss this possibility. 

On the other hand, in the split SUSY all the superpartners of the SM chiral fermions are assumed to be heavy
while the superpartners of bosonic particles in the SM, i.e. gauginos and Higgsinos remain relatively light.
One such idea realizing the split SUSY is to introduce the so-called $Z^\prime$ mediated SUSY breaking \cite{Zprime} or U(1)$^\prime$ mediation, 
in which a U(1)$^\prime$ gauge sector plays the role of the messenger sector for SUSY breaking in the visible sector.  
In U(1)$^\prime$ mediation, scalar components of chiral superfields charged under U(1)$^\prime$ acquire quite heavy masses of order $100~{\rm TeV}$ at one-loop level, 
while the MSSM gauginos get masses of $10^{2-3}~{\rm GeV}$ at two-loop level. 
Moreover, by employing a family dependent U(1)$^\prime$ charge assignment, one can achieve a hierarchical sfermion spectrum. 
Thus, e.g. the third family of squarks can be made the lightest \cite{JKS} or heaviest family \cite{Kim}. 
%

With the original forms of the effective SUSY and split SUSY, however, the discrepancy of
$(g-2)_\mu$ observed at BNL from the MSSM prediction cannot be accommodated.
In this Letter, we attempt to obtain a more desirable spectrum
on the superparticles by employing a proper  U(1)$^\prime$ mediation
such that the $(g-2)_\mu$ observed at BNL is explained in the SUSY framework.
To be consistent with the LHC results, the first two generations of squarks should be made
quite heavy while the second generation of sleptons and the stop need to remain light to
explain $(g-2)_\mu$ and to avoid the fine-tuning problem.
Indeed, such a hierarchical spectrum of the superparticles is hard to obtain in
the conventional SUSY  breaking scenarios. In order to obtain such a spectrum,
we need to find a desirable U(1)$'$ gauge symmetry useful for the U(1)$^\prime$ mediation.

It is well-known that with the MSSM {\it chiral} matter contents, U(1)$_B$ and U(1)$_L$ are anomalous
for each generation. However, their proper combinations could cancel all the gauge anomalies.
For instance, U(1)$_{B-L}$ is anomaly-free for each  generation.
U(1)$_{B_i-B_j}$ and U(1)$_{L_i-L_j}$ ($i\neq j$) are also anomaly-free, where $i$ and $j$ denote
generation numbers. Of course, all the linear combinations of U(1)$_{B-L}$, U(1)$_{B_i-B_j}$
and U(1)$_{L_i-L_j}$ are also  anomaly-free.
Hence, U(1)$_{B_i-L_j}$ ($i\neq j$) can be an anomaly-free gauge symmetry. Indeed,  U(1)$_{\pm(B_1+B_2-L_1-L_3)+B_3-L_3}$
was adopted for U(1)$^\prime$ mediation \cite{Kim} where the smuon/sneutrino can be light
enough for an explanation of $(g-2)_\mu$.

In this Letter, we propose a U(1)$^\prime$ mediation with 
\dis{
{\rm U(1)}_{B_1+B_2-2L_1} ,
}
which is a linear combination of U(1)$_{B_2-B_1}$ and U(1)$_{B_1-L_1}$.
By virtue of this gauge symmetry, proton stability can be protected:
U(1)$_{B_1+B_2-2L_1}$ disallows all the dimension 4 and 5 baryon number violating operators.\footnote{U(1)$_{B_1+B_2-L_1-L_3}$ is also a quite interesting gauge symmetry,
in which the stop and the smuon/sneutrino can still be light.
However, we note that the extra U(1) gauge symmetry allows a heavy right-handed neutrino mass term
in the superpotential only for one right-handed neutrino, unless they are broken at a high energy scale.
For a successful seesaw mechanism, we intend to cheaply obtain heavy mass terms for at least {\it two}
right-handed neutrinos, which should be neutral under the extra U(1). It is another motivation to
study U(1)$_{B_1+B_2-2L_1}$.}
Hence, the first two generations of squarks and the  selectron are expected to be quite heavy
through the U(1)$^\prime$ mediation mechanism since they carry nonzero charges of U(1)$_{B_1+B_2-2L_1}$.
On the other hand, the stops and smuons/sneutrino remain relatively light.
Their masses could be obtained through the ordinary gravity or gauge mediation mechanism.

The paper is organized as follows.
In Section II, we introduce the U(1)$_{B_1+B_2-2L_1}$ mediated SUSY breaking,
and attempt to obtain the low energy spectrum of superparticles.
We also discuss $(g-2)_\mu$. In Section III, we propose a concrete U(1)$'$ mediation model.
Section IV is a conclusion.

\section{superparticle spectrum and $(g-2)_\mu$}

We suppose that the visible and the hidden sectors can communicate through a U(1) gauge interaction.
The SUSY breaking in the hidden sector, which can be parametrized with a spurion field $X=M+\theta^2F$,
is assumed to generate a mass of the gaugino of  U(1)$^\prime$ ($\equiv M_{\tilde{Z}^\prime}$) at the scale $\Lambda_S$ \cite{Zprime}:
\dis{
M_{\tilde{Z}^\prime}\sim \frac{g_{Z^\prime}^2}{16\pi^2}\frac{F}{M}  ,
}
which plays the role of the order parameter of the SUSY breaking effects in the visible sector. 
%
%
The SUSY breaking in the U(1)$^\prime$ gauge sector
can be transferred to the visible sector \cite{Zprime},
inducing the soft masses for the scalar components of the superfields carrying U(1)$^\prime$ charges:
\dis{ \label{mf}
\widetilde{m}^2_{f}\sim\frac{Q_{f}^2g_{Z^\prime}^2}{16\pi^2}M_{\tilde{Z}^\prime}^2{\rm log}\left(\frac{\Lambda_S}{M_{\tilde{Z}^\prime}}\right) ,
}
where $Q_f$ stands for the U(1)$^\prime$ gauge charge.
We note here that it can be designed to solve the flavor changing neutral coupling (FCNC) problem in the gravity mediation scenario. 
By assigning the same U(1)$^\prime$ charges to the families needed to avoid the FCNC problem, we can get the degenerate and diagonal form of mass matrices for those families of superpartners.
If $\widetilde{m}_f^2$ dominates the soft mass squared over the one induced by gravity mediation,
it does not have to be much heavier than 1 TeV to avoid the FCNC.
Unlike the original U(1)$^\prime$ mediation of Ref.~\cite{Zprime}, thus,
we assume that the soft scalar masses generated through this mechanism are heavy enough just to avoid the experimental bounds on superparticles at the LHC.
For \eq{mf} we suppose that $M_{\tilde{Z}^\prime}$ is of order $10^{4}~{\rm GeV}$ and $\Lambda_S\sim 10^{16}~{\rm GeV}$.
Hence, for instance, $M$ would be of order $10^{14}~{\rm GeV}$ if $F\sim (10^{10}~{\rm GeV})^2$, and $M\sim 10^{10}~{\rm GeV}$ if $F\sim (10^8~{\rm GeV})^2$.

As seen in \eq{mf}, the RG running of $\widetilde{m}^2_{f}$ between the two energy scales $\Lambda_S$ and $M_{\tilde{Z}^\prime}$ is dominated by the U(1)$^\prime$ gaugino.
Below the $M_{\tilde{Z}^\prime}$ scale, however, the U(1)$^\prime$ gaugino is decoupled, and so the RG running of $\widetilde{m}^2_{f}$ is governed only by the RG equations of the relevant MSSM superfields and the U(1)$^\prime$ gauge field.
Accordingly, below the $M_{\tilde{Z}^\prime}$ scale the change of $\widetilde{m}^2_{f}$ would become much smaller.

The mass splitting between the bosonic and fermionic modes in the chiral matter sector as well as in the  U(1)$^\prime$ gauge sector generate also the MSSM gaugino masses at two-loop level \cite{Zprime}.
Hence, the MSSM gaugino masses are much suppressed compared to the soft masses of the scalars with U(1)$^\prime$ charges:
\dis{ \label{Ma}
\delta M_k\sim \frac{g_{Z^\prime}^2g_k^2}{(16\pi^2)^2}M_{\tilde{Z}^\prime}{\rm log}\left(\frac{\Lambda_S}{M_{\tilde{Z}^\prime}}\right)\sim 10~g_k^2 ~{\rm GeV} ,
}
where $g_k$ ($k=3,2,1$) denotes the MSSM gauge couplings. These gaugino masses are too light,
and so they are dominated by other mediation mechanisms of SUSY breaking so that they are of order $10^{2-3}~{\rm GeV}$.  We will neglect the contributions by \eq{Ma} to the MSSM gaugino masses.

%
%
\begin{table}[!h]
\begin{center}
\begin{tabular}
{c|cccccc|cccccc}
{\rm Superfields}  &   ~$q_{1,2}$~   &
 ~$u_{1,2}^c$~  &  ~$d_{1,2}^c$~  &  ~$l_{1}$~  &
 ~$\nu_{1}^c$~  &  ~$e_{1}^c$~ & ~~$N$~  & ~$N^c$~  & ~$N_H$~  & ~$N_H^{c}$~  & ~$N_H^\prime$~  & ~$N_H^{c\prime}$~
  \\
\hline
U(1)$_{B_1+B_2-2L_1}$ & ~$\frac13$ & $-\frac13$ & $-\frac13$ & $-2$ & ~$2$ & $-2$ & $-2$ & ~$2$ & ~$2$ & $-2$ & ~$\frac13$ & $-\frac13$
\\
U(1)$_{\rm PQ}$ & ~$\frac12$ & ~$\frac12$ & ~$\frac12$ & ~$\frac12$ & $-\frac12$ & ~$\frac12$ & ~$\frac32$ & ~$\frac12$ & $-1$ & $-1$ & $-2$ & ~$0$
\end{tabular}
\end{center}\caption{Matter fields carrying  U(1)$_{B_1+B_2-2L_1}$ {\it gauge} charges.
$\{N_{(H)}^{(\prime)},N_{(H)}^{c(\prime)}\}$ are MSSM singlets. 
The scalar components of them acquire
heavy masses of $7-40~{\rm TeV}$ depending on their charges (and the full matter contents)
through the U(1)$_{B_1+B_2-2L_1}$ mediated SUSY breaking mechanism.
}\label{tab:Qnumb1}
\end{table}

We identify U(1)$^\prime$ with U(1)$_{B_1+B_2-2L_1}$, which is anomaly-free.
The charge assignments of U(1)$_{B_1+B_2-2L_1}$ are presented in Table I.
From \eq{mf} the masses of the first and second generations of squarks are in the range of $5-10 ~{\rm TeV}$, and the masses of the first generation of the sleptons are of $\sim 30-60~{\rm TeV}$.
Since the third generation of the quarks, and the second and third generations of leptons are neutral under U(1)$_{B_1+B_2-2L_1}$, their superpartners get the soft masses only from the ordinary gravity or gauge mediation.
The MSSM gaugino masses are also dominated by those  mechanisms. Thus, their masses of order $10^{2-3}~{\rm GeV}$ are expected.

The masses of $5-10 ~{\rm TeV}$ for the first two squark families and $500~{\rm GeV}$ for the third squark family
satisfy the LHC mass bounds on the extra colored particles. Such a hierarchical mass spectrum  is non-trivial to realize in the constrained MSSM\,(CMSSM).
In our U(1)$^\prime$ mediation, a serious fine-tuning problem in the Higgs sector can be avoided due to the relatively light stop.
We will briefly discuss how to explain 126 GeV Higgs mass later.
The relatively light smuon/sneutrino and neutralino/chargino admit the possibility that $(g-2)_\mu$ deviated from the SM prediction is supported from SUSY, as will be discussed later.

The SUSY FCNC problems are associated with the ratios of the off-diagonal to the diagonal components in the soft mass matrices.
In the U(1)$^\prime$ mediation, where the  U(1)$^\prime$ charges are given by $B_1+B_2-2L_1$,  SUSY breaking effects generates {\it degenerate} heavy soft masses
($5-10~{\rm TeV}$) in the diagonal $(11)$ and $(22)$ components in the squark mass matrix.
Since the masses induced by the U(1)$^\prime$ mediation dominate gravity mediation for the soft masses,
the FCNC in the quark sector can be sufficiently suppressed.
Also, the U(1)$^\prime$ mediation induces very heavy masses ($\sim 40~{\rm TeV}$) to the first family of sleptons.
Note that the U(1)$_{B_1+B_2-2L_1}$ symmetry forbids the off-diagonal elements $M^2_{\rm lepton\,(1i)}\,(i=2,3)$  in the slepton mass matrix. 
If U(1)$_{B_1+B_2-2L_1}$ survives down to low energies, thus, the FCNC problem in the lepton sector, 
e.g.  such as $\mu^-\rightarrow e^-\gamma$ and $\tau^-\rightarrow e^-\gamma$, could not be severe.\footnote{We suppose that the $(2,3)$ and $(3,2)$ elements in the slepton mass matrix are small enough for suppression of $\tau^-\rightarrow\mu^-\gamma$. 
Note, however, that the bound on the $\tau^-$ decay is less severe, BR$(\tau^-\rightarrow \mu^-\gamma)<4.4\times 10^{-8}$ \cite{PDG}. 
}
%
%

In supergravity models, the predicted mass relation for the MSSM gauginos is
\dis{ \label{gauginoMass}
\frac{M_3(Q)}{g_3^2(Q)}=\frac{M_2(
Q)}{g_2^2(Q)}=\frac{M_1(Q)}{g_1^2(Q)}=\frac{m_{1/2}}{g_U^2} ,
}
where $Q$ denotes the renormalization energy scale, $m_{1/2}$ and $g_U$ mean the gaugino mass and gauge coupling at the GUT scale.
In \eq{gauginoMass}, the MSSM gauge couplings are given by
\begin{eqnarray} \label{g_k}
g_k^2(t)=\frac{g_U^2}
{1-\frac{g_U^2}{8\pi^2}b_k(t-t_0)} \qquad {\rm for} ~~k=3,~2,~1,
\end{eqnarray}
where $t-t_0\equiv {\rm log}(Q/M_{\rm GUT})$, and  $b_k$ ($k=3,2,1$) denotes the beta function coefficients of the gauge couplings for SU(3)$_c$, SU(2)$_L$ and U(1)$_Y$.
If there exist extra $v$ pairs of $\{{\bf 5},\overline{\bf 5}\}$, $b_k$ are given by $b_k=(-3+v,1+v,33/5+v)$.
The unified gauge coupling $g_U^2$ is estimated as $0.52$, $0.62$, $0.82$, and $1.18$, for $v=0,1,2,$ and 3, respectively.
Note that SU(3)$_c$ becomes (approximately) conformal for $v=3$,
because $b_3$ vanishes at one-loop level.
Since the low energy magnitudes of the MSSM gauge couplings are related to each other
as $g_3^2:g_2^2:g_1^2\approx 6:2:1$,
we have $M_3:M_2:M_1\approx 6:2:1$ at the TeV scale.
For instance, thus, we have $M_2\approx 400~{\rm GeV}$ and $M_1\approx 200~{\rm GeV}$ for $M_3\approx  1.2~{\rm TeV}$.
In the CMSSM ($v=0$), it can be achieved by setting $m_{1/2}\approx 532~{\rm GeV}$ at the GUT scale.
On the other hand, for $v=1$ or $3$,
which are the cases of our particular interest in this Letter,
$m_{1/2}$ should be taken as a larger value, 631 GeV ($v=1$) or 1.2 TeV ($v=3$) in order to get $M_3=1.2~{\rm TeV}$ at the TeV scale.
In the next section, we will propose a concrete model with $v=3$.

We require $(500~{\rm GeV})^2\lesssim \widetilde{m}_t^2\ll (1~{\rm TeV})^2$,
which can be obtained from $m_0^2\lesssim (100~{\rm GeV})^2$ at the GUT scale in the CMSSM ($v=0$), if $m_{1/2}\approx 300~{\rm GeV}$.
In this case, the slepton masses of the second generation become much lighter than $500~{\rm GeV}$.
In the case $m_{1/2}\gtrsim 1~{\rm TeV}$, however,
the stop mass squared rapidly decreases with energy, and becomes negative at some high energy scale.
It means that the stop mass should be much heavier than 500 GeV, if it is taken to be positive at the GUT scale.
We will see that such a heavy gluino effect on the RG equation could be compensated with the two-loop effects by very heavy superpartners,
maintaining the positive squared mass of the stop all the way up to the GUT scale.

The relatively light smuon/sneutrino (the second generation of the sleptons) and neutralino/chargino could be
responsible for the deviation of the muon $g-2$ from the SM prediction \cite{MSSMg-2,Kim,jp,EHIY}.
In the MSSM, the muon $g-2$ is
contributed by the sneutrino/chargino loops and the smuon/neutralino loops.
For a large $\mu$, i.e. $\mu^2\gg \widetilde{m}_{\mu L,R}^2$, the contribution by the smuon/neutralino loops dominates that by the the sneutrino/chargino loops \cite{MSSMg-2,Kim}. 
If $M_2,\mu\sim 500~{\rm GeV}$, and the sleptons' masses are quite smaller than this value, i.e. $\ll 500~{\rm GeV}$, however,
the sneutrino/chargino loop becomes dominant \cite{MSSMg-2,EHIY}:
\dis{
\Delta (g-2)_\mu\approx \frac{g_2^2 m_\mu^2}{16\pi^2 M_2\mu}~{\rm tan}\beta~I^+(x,y)\approx 1.9\times 10^{-9}\left(\frac{{\rm tan}\beta}{20}\right)\left\{\frac{(400~{\rm GeV})^2}{M_2\mu}\right\}I^+(x,y) ,
}
where $m_\mu$ is the muon mass.
The loop function $I^+(x,y)$ is defined as
\dis{
I^+(x,y)\equiv xy\left[\frac{13-7(x+y)+xy}{(x-1)^2(y-1)^2}
-\frac{(4+2x){\rm log}x}{(x-y)(x-1)^3}
+\frac{(4+2y){\rm log}y}{(x-y)(y-1)^3}\right] ,
}
where $x\equiv M_2^2/\widetilde{m}_{\mu L}^2$ and $y\equiv \mu^2/\widetilde{m}_{\mu L}^2$.
Here, we assume $\widetilde{m}_{\mu L}^2,\widetilde{m}_{\nu L}^2\gg M_W^2$ and set $\widetilde{m}_{\mu L}^2\approx \widetilde{m}_{\nu L}^2$.
For $2<x,y<5$, $I^+(x,y)$ is in the range of  $1.0-1.1$.
Thus, $\Delta(g-2)_\mu\approx 2.6\times 10^{-9}$ can be achieved for a large ${\rm tan}\beta$ [$\gtrsim 20$] and a small $M_2\mu$ [$\lesssim (400~{\rm GeV})^2$]. 
Since an excessively large $\mu$ could make the fine-tuning in the Higgs sector more serious, we take the  case that the sneutrino/chargino loop is dominant. 

Now let us explore the conditions for $\widetilde{m}_{t}^2\approx (500~{\rm GeV})^2$ and $\widetilde{m}_{\mu}^2\approx (400~{\rm GeV})^2$ at the EW scale. 
First, we consider the RG behavior of the heavier soft mass squareds, namely, the mass squareds of the first two generations of squarks and the first generation of sleptons, which are charged under U(1)$_{B_1+B_2-2L_1}$.
The RG evolution of such heavy soft mass squareds are mainly governed by the U(1)$_{B_1+B_2-2L_1}$ gaugino, because it is much heavier ($\sim 10^4~{\rm GeV}$) than other superpartners: 
\dis{ \label{RGZprime}
16\pi^2\frac{d\widetilde{m}_f^2}{dt}\approx -8Q_f^2g_{Z^\prime}^2(t)|M_{\tilde{Z}^\prime}(t)|^2 . 
}
Here, $g_{Z^\prime}^2(t)$ and $M_{\tilde{Z}^\prime}(t)$ 
are given by $g_{Z^\prime 0}^2/[1-\frac{g_{Z^\prime 0}^2}{8\pi^2}b_{Z^\prime}(t-t_0)]$ and $M_{\tilde{Z}^\prime 0}/[1-\frac{g_{Z^\prime 0}^2}{8\pi^2}b_{Z^\prime}(t-t_0)]$, respectively, 
where $g_{Z^\prime 0}^2$ and $M_{\tilde{Z}^\prime 0}$ indicate the boundary values of $g_{Z^\prime}^2(t)$ and $M_{\tilde{Z}^\prime}(t)$ at the GUT scale, 
and $b_{Z^\prime}$ is the beta function coefficient of the U(1)$_{B_1+B_2-2L_1}$ gauge coupling.  
\eq{RGZprime} is integrable. The analytic solution of \eq{RGZprime} is 
\dis{ \label{heavySol}
\widetilde{m}_f^2(t) \approx \frac{2Q_f^2}{b_{Z^\prime}}M_{\tilde{Z}^\prime 0}^2\left[1-\frac{g_{Z^\prime}^4(t)}{g_{Z^\prime 0}^4}\right]\equiv Q_f^2m_0^2 \left[1-\frac{g_{Z^\prime}^4(t)}{g_{Z^\prime 0}^4}\right] .
}
For a large $b_{Z^\prime}$, $g_{Z^\prime}^2(t)$ rapidly drops down, and so $\widetilde{m}_f^2(t)$ quickly approaches a constant ($=Q_f^2m_0^2$) as in the case considered in Ref.~\cite{tachyon}. 
In our case, $b_{Z^\prime}=\frac{314}{9}$ by the charge assignment in Table I.  
Thus, we will ignore the second term of \eq{heavySol}.  

Next, we discuss the RG behavior of the lighter soft mass squareds, namely, the third generation of squarks, and the second (and third) generation(s) of sleptons.   
We require $M_2\approx 400~{\rm GeV}$ at the EW scale for explaining $(g-2)_\mu$.
Accordingly, the gluino mass should be $1.2~{\rm TeV}$ at low energy, as discussed above.
Since the mass of gluino [$M_3(t)=m_{1/2}g_3^2(t)/g_U^2$] as well as the masses of the first two generations of squarks [$\widetilde{m}^2_{q_{1,2}}(t)\approx(\frac13)^2m_0^2$] 
and the first generation of sleptons [$\widetilde{m}^2_{l_{1}}(t)\approx 2^2m_0^2$] are assumed to be quite heavy, 
their contributions to the RG equations are dominant over the Yukawa couplings' contributions at low energies.
Neglecting the U(1)$_Y$ and Yukawa couplings' contributions, the RG equations for $\widetilde{m}_{t}^2(t)$ and $\widetilde{m}_{\mu}^2(t)$ are approximately given by
\dis{ \label{RGeq}
&\frac{d\widetilde{m}_{t}^2}{dt}\approx -\frac{32}{3}\frac{g_3^2}{16\pi^2}|M_3|^2
+\left\{\frac{128}{3}\frac{g_3^4Q_q^2}{(16\pi^2)^2}+\frac{g_2^4}{(16\pi^2)^2}\left(18Q_q^2+3Q_l^2\right)\right\}m_0^2 ,
\\
&\qquad\qquad~~ \frac{d\widetilde{m}_{\mu}^2}{dt}\approx -6\frac{g_2^2}{16\pi^2}|M_2|^2
+\frac{g_2^4}{(16\pi^2)^2}\left(18Q_q^2+3Q_l^2\right)m_0^2 ,
}
where the terms with $Q_{q,l}^2$ correspond to the two-loop effects by the heavy superpartners \cite{tachyon}.
$Q_{q}$ ($=\pm\frac13$) [$Q_l$ ($=\mp 2$)] denotes the
U(1)$_{B_1+B_2-2L_1}$ charge for the heavy  generation(s) of the squarks [sleptons].
Thus, the two generations of (s)quarks and one generation of (s)leptons make contributions to the two-loop effects in \eq{RGeq}.
Here, we inserted the expression $\widetilde{m}^2_f(\mu)\approx Q_f^2m_0^2$ in \eq{heavySol}.
The above approximations are possible because $m_0^2,M_3^2\gg \widetilde{m}^2_t$, and $g_3^2$ is of order unity at low energies.

\begin{figure}
\begin{center}
\subfigure[]
{\includegraphics[width=0.48\linewidth]{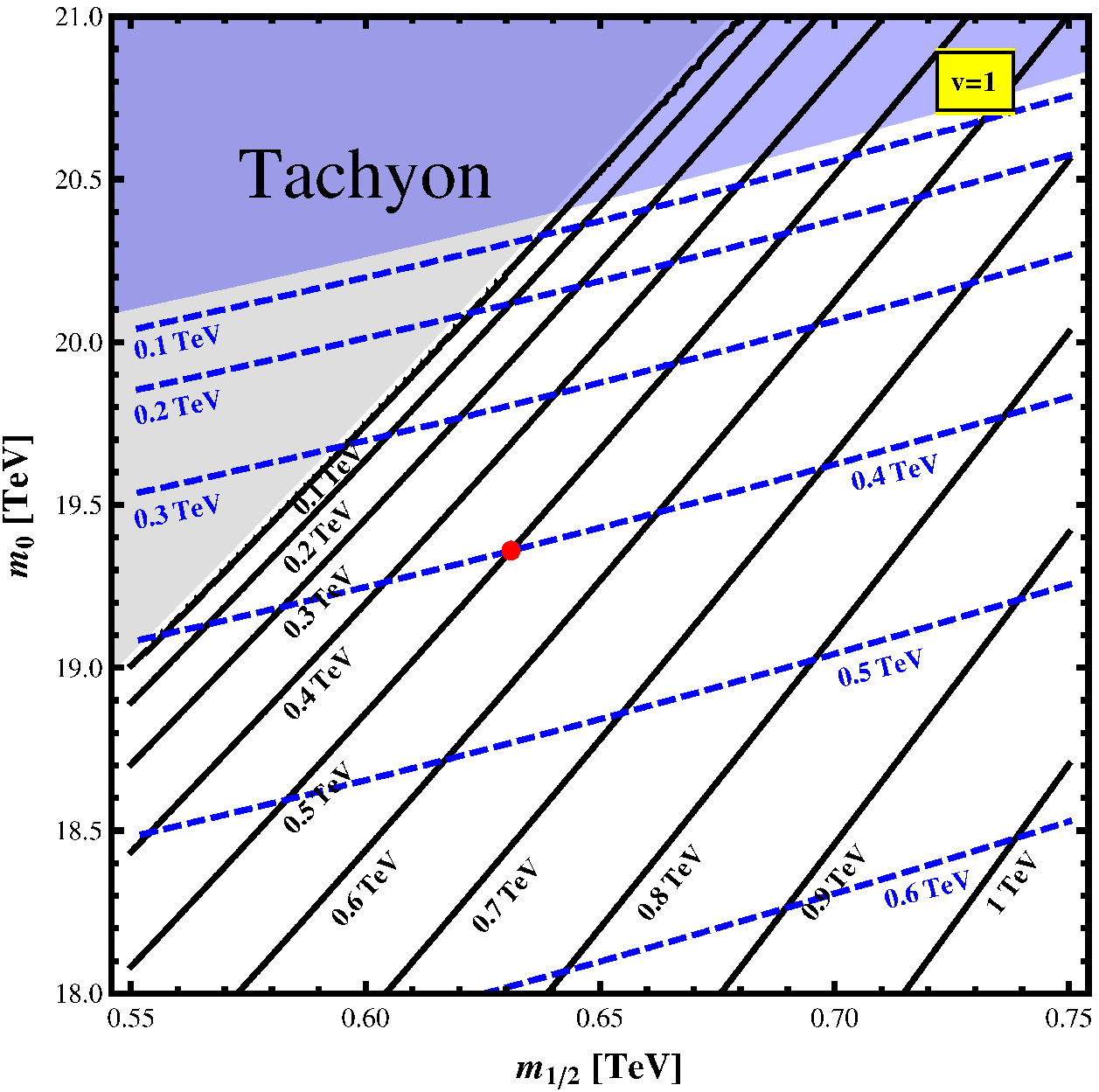}}
\hspace{0.2cm}
\subfigure[] 
{\includegraphics[width=0.48\linewidth]{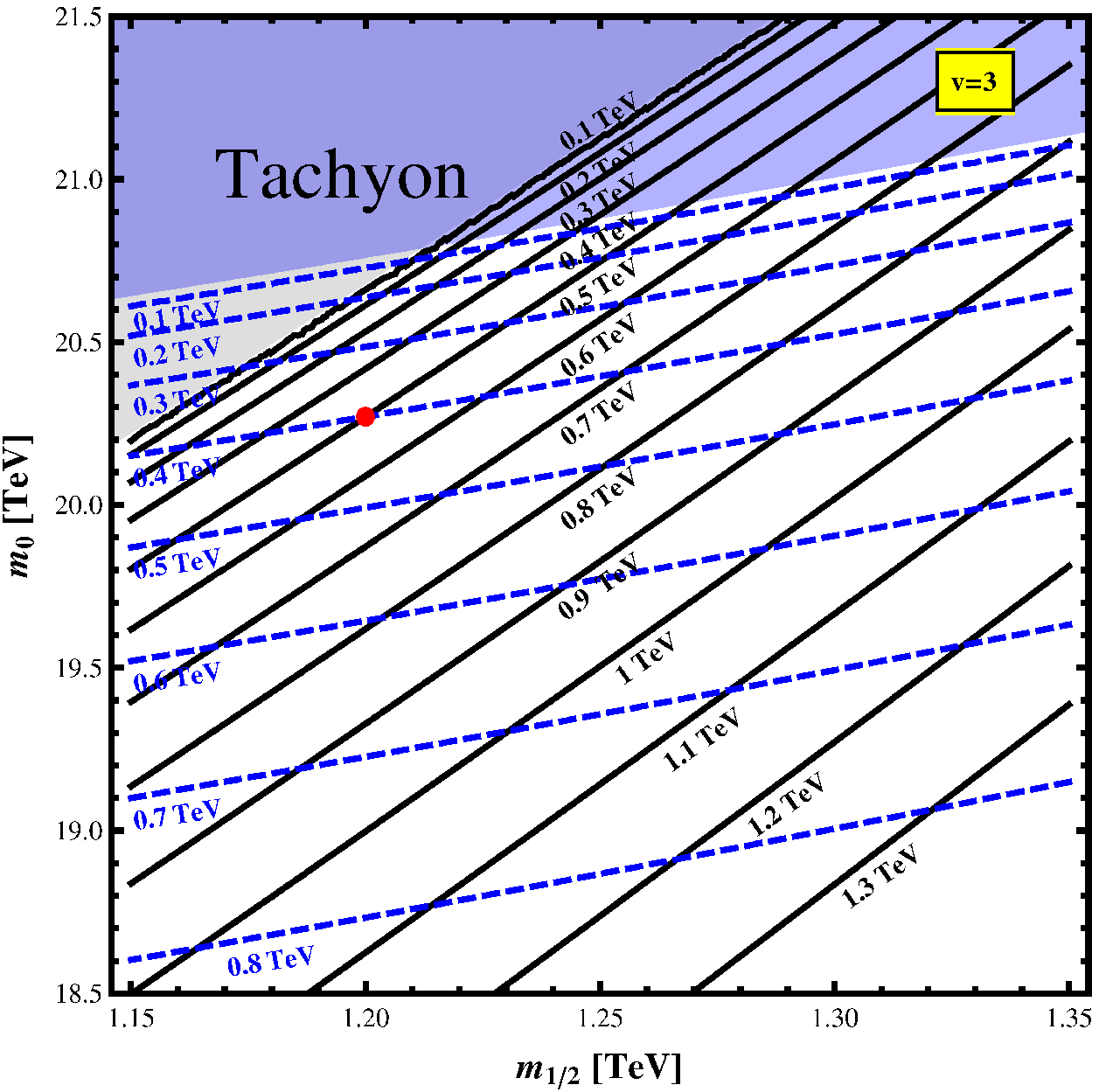}}
\end{center}
\caption{Low energy values of $(\widetilde{m}_t,\widetilde{m}_\mu)$ vs. $(m_{1/2},m_0)$ for $v=1$ (a) 
and $v=3$ (b). The solid [dotted] lines are contour lines of the same masses of $\widetilde{m}_t$ [$\widetilde{m}_\mu$] for various $(m_{1/2}, m_0)$. 
$\widetilde{m}_0^2$ is taken as $(1.2~{\rm TeV})^2$ 
in (a) and $(1.7~{\rm TeV})^2$ in (b), respectively.  
$m_{1/2}\approx 0.63~{\rm TeV}$ in (a) and $m_{1/2}\approx 1.20~{\rm TeV}$ in (b) provide $(M_3, M_2, M_1)\approx (1.2~{\rm TeV}, ~0.4~{\rm TeV},~ 0.2~{\rm TeV})$ at TeV scale.  
} \label{fig:m0m12}
\end{figure}

Using Eqs.~(\ref{gauginoMass}) and (\ref{g_k}), the above equations are integrable. Requiring
$\widetilde{m}_{t}^2\approx (500~{\rm GeV})^2$  $\widetilde{m}_{\mu}^2\approx (400~{\rm GeV})^2$, and
$M_3\approx 1.2~{\rm TeV}$ at TeV scale, thus,
we have
\dis{ \label{RGsol}
&(500~{\rm GeV})^2\approx \widetilde{m}_0^2+C^q_1 m_{1/2}^2-C^q_2m_0^2 ,
\\
&(400~{\rm GeV})^2\approx \widetilde{m}_0^2+C^l_1 m_{1/2}^2-C^l_2m_0^2 .
}
Here $\widetilde{m}_0^2$ indicates a common boundary value of $\widetilde{m}_{t}^2(t)$ and  $\widetilde{m}_{\mu}^2(t)$ at the GUT scale,
which could be generated e.g. through the gravity mediated SUSY breaking mechanism.
As discussed above, the boundary value of the gluino mass, $m_{1/2}$ is given by 631 GeV for $v=1$ and 1.2 TeV for $v=3$.
The coefficients in \eq{RGsol} are roughly estimated as $[C^q_1,C^q_2;C^l_1,C^l_2]\approx[3.50, 0.007; 0.45, 0.004]$ and $[2.39, 0.015; 0.33, 0.008]$ for $v=1$ and $3$, respectively.
They yield $[\widetilde{m}_0^2,m_0^2]\approx [(1.2~{\rm TeV})^2,(19.4~{\rm TeV})^2]$ for $v=1$, and $[(1.7~{\rm TeV})^2, (20.3~{\rm TeV})^2]$ for $v=3$.
They imply that the squared masses for the first two generations of squarks and the first generation of sleptons are given by $[\widetilde{m}^2_{q_{1,2}}, \widetilde{m}^2_{l_{1}}]\approx [(6.5~{\rm TeV})^2, (38.7~{\rm TeV})^2]$ for $v=1$, and $[(6.8~{\rm TeV})^2, (40.5~{\rm TeV})^2]$ for $v=3$ at TeV scale.
For the low energy values of $(\widetilde{m}_t, \widetilde{m}_\mu)$ resulted from other boundary  values of $(m_{1/2}, m_0)$, see Fig. \ref{fig:m0m12}. 
Since the contour lines in the $v=3$ case are denser than those in the $v=1$ case, $\widetilde{m}_t$ and $\widetilde{m}_\mu$ are more sensitive to the choices of $m_{1/2}$ and $m_0$ in the $v=3$ case. 

So far we have ignored the Yukawa couplings.
They might be important around the GUT scale, 
since the $m_0^2$ terms in \eq{RGeq} or $\widetilde{m}_f^2(t)$ in \eq{heavySol} vanish at the boundary.
However, even the top quark Yukawa coupling decreases with energy, eventually down to about 0.5 (0.2) for ${\rm tan}\beta=20$ and $v=1$ ($v=3$) at the GUT scale.
Thus, the $\widetilde{m}_0^2$ ($m_0^2$) turns out to be just a bit larger (smaller) than the above estimation, when the Yukawa couplings are also considered in the RG equations.

$[\widetilde{m}_0^2,m_0^2]$ chosen in the above might determine also the soft mass squareds of the Higgs,  $\widetilde{m}_{h_u}^2$ and $\widetilde{m}_{h_d}^2$, 
and so affect the $Z$ boson mass at the minimum of the Higgs scalar potential. 
However, a mechanism introduced in the next section for explaining 126 GeV Higgs mass (with vector-like leptons $\{\hat{L},\hat{L}^c;\hat{N},\hat{N}^c\}$ \cite{KS2}) are also involved there. 
Another free parameter, $\mu$ should also be considered. 
Moreover, one can take other boundary boundary values different from $\widetilde{m}_0^2$ for $\widetilde{m}_{h_u}^2(t)$ and $\widetilde{m}_{h_d}^2(t)$. 
In this Letter, we don't discuss their RG behaviors, 
and the EW symmetry breaking in details.

\section{The Model}

Unlike the first and second generations of (s)quarks and the first generation of (s)leptons, the other MSSM superfields do not carry  U(1)$_{B_1+B_2-2L_1}$ gauge charges.
Hence the mixing between the first (last) two and the third (first) generations in the quark (lepton) sector is impossible, if U(1)$_{B_1+B_2-2L_1}$ remains unbroken. 
In this section, we will describe how the desired mixing in the quark and lepton sectors can be induced.

We introduce a global U(1)$_{\rm PQ}$ symmetry.
The global charges for the superfields neutral under U(1)$_{B_1+B_2-2L_1}$ are listed in Table II.
The renormalizable superpotential consistent with the gauged U(1)$_{B_1+B_2-2L_1}$ and the global U(1)$_{\rm PQ}$ symmetries is 
\dis{ \label{W}
W_{\rm }=&\sum_{i,j=1,2}\left(y^{ij}_{u}q_ih_uu_j^c +y_{d}^{ij}q_ih_dd_j^c\right)
+\sum_{i,j=2,3}\left(y_{\nu}^{ij}l_ih_u\nu_j^c+y_{e}^{ij}l_ih_de_j^c
+\frac12M^{ij}\nu_i^c\nu_j^c\right)
\\
&\qquad\qquad +y_{t}q_3h_uu_3^c+y_{b}q_3h_dd_3^c
+y_{e}l_1h_de_1^c 
+\mu h_uh_d ,
}
where the ``$y$''s denote the dimensionless Yukawa coupling constants, and ``$M^{ij}$'' ($=M^{ji}$) means the Majorana masses for the two right-handed neutrinos neutral under U(1)$_{B_1+B_2-2L_1}$.
The Majorana mass terms of the right-handed neutrinos
and the MSSM $\mu$ term break U(1)$_{\rm PQ}$ explicitly. We will explain later how they can be generated with the desired sizes.
Note that there are no Dirac and Majorana mass terms for $\nu_1^c$.
Only with the two heavy right-handed neutrinos, 
however, the seesaw mechanism and leptogenesis (through the CP violating phase of heavy neutrinos) are still possible \cite{2seesaw}.

%
%
\begin{table}[!h]
\begin{center}
\begin{tabular}
{c|ccccccc|cccccc}
{\rm Superfields}  &    ~$q_3$~  & ~$u_3^c$~ & ~$d_3^c$~ & ~$l_{2,3}$~ & ~$\nu_{2,3}^c$~ & ~$e_{2,3}^c$~ & ~$h_{u,d}$~ 
& ~$D$~  & ~$D^c$~ 
& ~$L$~  & ~$L^c$~  
& ~$P$~ & ~$Q$~
  \\
\hline
U(1)$_{\rm PQ}$ &  ~$\frac12$ & ~$\frac12$ & ~$\frac12$ & ~$\frac32$ & $-\frac12$ & $-\frac12$ & $-1$ 
& ~$\frac32$ & ~$\frac12$  
& ~$1$ & ~$1$  
& ~$1$ & $-1$
%
\end{tabular}
\end{center}\caption{Matter fields {\it neutral} under the U(1)$_{B_1+B_2-2L_1}$ gauge symmetry.
$\{D,D^c; L,L^c\}$ compose the $\{{\bf 5},\overline{\bf 5}\}$ of SU(5), 
while $\{P,Q\}$ carry only the global U(1)$_{\rm PQ}$ charges.
The scalar components of them get relatively lighter masses of order $100~{\rm GeV}$ only from ordinary gravity or gauge mediated SUSY breaking.
}\label{tab:Qnumb1}
\end{table}

In order to break U(1)$_{B_1+B_2-2L_1}$, we introduce the MSSM singlets $\{N_H,N_H^c;N_H^\prime,N_H^{c\prime}\}$, whose gauge and global quantum numbers are displayed in Table I.
They can develop vacuum expectation values (VEVs), breaking U(1)$_{B_1+B_2-2L_1}$, as will be explained below.
For successful mixing in the lepton and quark sectors, we need the MSSM singlets and color triplets, $\{N,N^c;D,D^c\}$,
where $D^c$ is assumed to have the same SM gauge quantum numbers with $d^c_i$.
For the gauge coupling unification, vector-like lepton doublets, $\{L,L^c\}$ need to be supplemented with a mass term of order EW scale.
Their gauge and global quantum numbers are presented in Tables I and II. 
Then, the following superpotential is also allowed:
\dis{ \label{Wmix}
&\quad ~~~~ W_{\rm mix}=\left(y_{\nu}^i N_H\nu^c_iN + \mu_NNN^c + y_{l}N^ch_ul_1\right)
+\left(y_{d}^jN_H^{\prime}d_j^cD+\mu_DDD^c
+y_{q}D^ch_dq_3\right)
\\
&\qquad\qquad\qquad\qquad\qquad ~
+\mu_{N_H}N_HN_H^{c} + \mu_{N_H}^\prime N_H^{\prime}N_H^{c\prime} + \mu_LLL^c ,
}
where $i=2,3$ and $j=1,2$. 
The sizable Yuakawa couplings $y_{\nu}^i$ and $y_{d}^j$ can drive the soft mass squareds of $\widetilde{N}_H$ and $\widetilde{N}_H^{\prime}$ negative at the EW scale.
Then the TeV scale VEVs, $\langle\widetilde{N}_H\rangle$ and $\langle\widetilde{N}_H^{\prime}\rangle$ can be generated, breaking U(1)$_{B_1+B_2-2L_1}$ completely. 
After decoupling the $\{N,N^c\}$ and $\{D,D^c\}$ below the $\mu_N$ or $\mu_D$ scale,
one can get the mixing terms of 
the {\it Dirac} neutrinos and d-type quarks,
whose effective Yukawa couplings are estimated as  
$y_{\nu}^iy_{l}\langle \widetilde{N}_H\rangle/\mu_N$, and $y_d^jy_q\langle\widetilde{N}_H^\prime\rangle/\mu_D$, respectively.
They fill the $(i,1)$ components of the mass matrices of Dirac neutrinos ($\equiv M_{D\nu}$), and the $(j,3)$ of the d-type quarks' mass matrix  ($\equiv M_d$), respectively.
The mass matrix for observed light neutrinos, $[M_\nu]^{pq}$ ($=[M_\nu]^{qp}=-[M_{D\nu}^TM^{-1}M_{D\nu}]^{pq}$, where $p,q=1,2,3$) can be obtained after integrating out $\{N,N^c\}$ and $\nu_i^c$.
%
%
The matrices diagonalizing $M_\nu$ 
and $M_d^\dagger M_d$, $U_L^{(\nu)}$ and $U_L^{(d)}$ are general enough to accommodate 
the PMNS and CKM matrices, respectively.  
Note that in contrast to the neutrinos and d-type quarks, the mass matrices of charged leptons and u-type quarks still remain block-diagonal.

The U(1)$_{\rm PQ}$ breaking terms of $\mu_{L,D}$, $\mu_{N_{(H)}}^{(\prime)}$ in \eq{Wmix} as well as $\mu$ in \eq{W} can be replaced by the nonrenormalizable terms  
with $P^2/M_P$ or $Q^{2}/M_P$ \cite{KimNilles84}, which respect the U(1)$_{\rm PQ}$ symmetry (for a recent discussion, see Ref.~\cite{Kim13Axion}). Here $M_P$ denotes the reduced Planck mass ($\approx 2.4\times 10^{18}~{\rm GeV}$).
The superfields $P$ and $Q$ get VEVs of order $10^{10}~{\rm GeV}$,
breaking the U(1)$_{\rm PQ}$ into the discrete $Z_2$ symmetry, e.g. through the superpotential $W\supset \kappa \Sigma(PQ-M_I^2)$ \cite{KimWglobalSym},
where $\Sigma$ is a superfield, $\kappa$ a dimensionless coupling, and $M_I$ indicates a mass parameter of order $10^{10}~{\rm GeV}$.\footnote{The presence of superheavy lepton doublets $\{L_G,L_G^c\}$ carrying $\mp 2$ [$\pm \frac32$] charges of U(1)$_{B_1+B_2-2L_1}$ [U(1)$_{\rm PQ}$] 
could open the possibility that
an extremely small Dirac neutrino mass ($=y_1y_2\langle P\rangle\langle h_u\rangle/M_G$) is  naturally generated e.g. from the superpotential, 
$W\supset y_1Pl_1L_G^c + M_GL_GL_G^c + y_2Lh_u\nu_1^c$, 
where $y_{1,2}\sim 10^{-3}$ and 
$M_G\sim 10^{16}~{\rm GeV}$.   
}  
Thus, $\langle P^2\rangle/M_P$ and $\langle Q^2\rangle/M_P$ can be of order $10^{2-3}~{\rm GeV}$.
The remaining $Z_2$ symmetry is identified with the matter parity in the MSSM, which forbids the $R$-parity violating terms.  
The Majorana masses $M^{ij}$ in \eq{W} also can be replaced by $P$.

The first and the other generations of {\it s}leptons are not mixed even in the mass eigen basis, because their masses are quite hierarchical [$(40~{\rm TeV})^2$ -- $({\cal O}(0.1)~{\rm TeV})^2$] and the U(1)$_{B_1+B_2-2L_1}$ breaking scale is low enough in this model.
Moreover, the mass matrix of the charged leptons also remains block-diagonal, as mentioned above.
Accordingly, $\mu^-\to e^-\gamma$ cannot arise 
through mediation by superparticles. 
The relatively light $(\tilde{\nu}_{\mu L}, \tilde{\mu}_{L})$ and $\tilde{\nu}_{\mu R}$ can be responsible for the deviation of $(g-2)_\mu$, as discussed above.  
Large mixing among the left-handed neutrinos doesn't  affect it.

Since the charge assignment of U(1)$_{B_1+B_2-2L_1}$ is not universal, the U(1)$_{B_1+B_2-2L_1}$ gauge boson could give rise to lepton flavor violations.  
As seen in \eq{heavySol}, however, the gauge coupling $g_{Z^\prime}$ is quite small at low energy, because U(1)$_{B_1+B_2-2L_1}$ survives down to low energy.   
%
%
Other FCNC effects by such a gauge boson 
can also be adequately suppressed, if the U(1)$_{B_1+B_2-2L_1}$ breaking scale is above a few TeV \cite{KS1}.

In the squark mass matrix, the diagonal components, $(1,1)$ and $(2,2)$ are degenerate with a squared mass of $(7~ {\rm TeV})^2$, while the off-diagonal components, $(j,3)$ and $(3,j)$, where $j=1,2$, remain zero due to U(1)$_{B_1+B_2-2L_1}$. 
The $(1,2)$, $(2,1)$, and $(3,3)$ can be filled dominantly by the gravity mediation effect, which are quite suppressed compared to the $(1,1)$ and $(2,2)$ components. 
After diagonalization in the fermionic quarks sector, $(1,2)$, $(2,1)$, and $(j,3)$, $(3,j)$ can be induced also by the mixing effect. 
The $(1,2)$ and $(2,1)$ components affect e.g. $K$-$\bar K$ mixing.   
The amplitude of $K$-$\bar K$ mixing by the squark mixing is roughly estimated as \cite{book}
\dis{ \label{KKbar}
{\cal M}_{K\bar{K}}\approx
\frac{4\alpha_3^2}{\widetilde m_q^2}
\left(\frac{\Delta\widetilde m_q^2}{\widetilde m_q^2}\right)^2 ,
}
where $\widetilde m_q^2\approx (7~{\rm TeV})^2$, and  $\Delta\widetilde m_q^2$ denotes the off-diagonal component of the squark mass matrix.
Since the SM still explains well the observed data, 
\eq{KKbar} should be smaller than the SM prediction,
${\cal M}_{K\bar{K}}^{\rm SM}\approx\alpha_2^2\sin^2\theta_c\cos^2\theta_c
(m_c^2/M_W^4)$, where $\theta_c$ stands for the Cabibbo mixing angle. 
The condition ${\cal M}_{K\bar{K}}\ll{\cal M}_{K\bar{K}}^{\rm SM}$ yields
\dis{
\left(\frac{\Delta \widetilde m_q^2}{\widetilde m_q^2}\right)\ll 5.6\times 10^{-2}\times 
\left(\frac{\widetilde m_q}{7~{\rm TeV}}\right) .
}
Note that when $\widetilde m_q=500~{\rm GeV}$, this estimation on $\Delta \widetilde m_q^2/\widetilde m_q^2$ ($\approx \delta_{12}^d$) provides slightly stronger constraint ($\ll 4\times 10^{-3}$) than those of Ref.~\cite{FV}. 
In order to suppress the squark mixing effect, hence, the off-diagonal element $(1,2)$ of the squark mass matrix should be much smaller than 5.6 percent of the diagonal one [i.e. $\ll (1.6~{\rm TeV})^2]$ when $\widetilde m_q=7~{\rm TeV}$.
If the mixings among the d-type quarks are given fully by the CKM (or a similar order mixing matrix) and the elements induced by the gravity mediation is smaller than $(1~{\rm TeV})^2$, this constraint can be satisfied. 
 
Finally, let us discuss the Higgs sector.
The simplest way to raise the Higgs mass in the framework of the MSSM is to consider the large $A$-term \cite{maximalmixing,text}, assuming $(A_t-\mu {\rm cot}\beta)^2/\widetilde{m}_t\approx 6$,
where $A_t$ indicates the $A$-term coefficient corresponding to the top quark Yukawa coupling.
However, the relation $(A_t-\mu {\rm cot}\beta)^2/\widetilde{m}_t^2\approx 6$ would be a fine-tuning relation.
Even in this case, we need one pair of $\{{\bf 5},\overline{\bf 5}\}$ (i.e. $v=1$) to induce the full mixing
among the three generations of the SM chiral fermions as discussed above.

Actually, the Higgs mass could easily be raised at tree level by promoting the MSSM $\mu$ term to the renormalizable superpotential $\lambda Sh_uh_d$, $\grave{a}$ $la$ the next to MSSM,
introducing a new singlet $S$ together with a new dimensionless coupling $\lambda$.
In fact, this approach was taken in the original suggestion of the U(1)$^\prime$ mediation  \cite{Zprime}.
For maintaining the perturbativity of the model all the way up to the GUT scale, however, $\lambda$ and ${\rm tan}\beta$ should be small enough: $0.6\lesssim \lambda\lesssim 0.7$ and $1\lesssim {\rm tan}\beta\lesssim 3$ \cite{LPnmssm}.
As seen in the above, however, a large ${\rm tan}\beta$ is needed to explain the central value of the observed
$(g-2)_\mu$.\footnote{The error is still $3.3\,\sigma-3.6\,\sigma$, and it is premature to exclude all the NMSSM models, not accommodating the BNL $(g-2)_\mu$.}

In this Letter, we will consider another method to raise the Higgs mass.
As pointed out in Ref.~\cite{KS2}, introduction of new  vector-like leptons $\{\hat{L},\hat{L}^{c};\hat{N},\hat{N}^{c}\}$ together with a new gauge symmetry is very helpful for raising the {\it radiative} Higgs mass,
where the new vector-like leptons are charged under a  new gauge symmetry: {\it all the superfields} (except $\{L,L^c\}$) {\it in Tables I and II are regarded as being neutral}.\footnote{In principle, $\{L,L^c\}$ can be identified with $\{\hat L,\hat L^c\}$.}
Due to the new gauge symmetry, we can take a relatively large Yukawa couplings between the new leptons and the Higgs field, avoiding a blowup of the Yukawa coupling below the GUT scale.
Of course one can identify the new gauge symmetry with U(1)$_{B_1+B_2-2L_1}$.
In this case, however, the charges of the new vector-like leptons should be small enough, say $\pm\frac{1}{40}$ (rather than $\mp 2$) so that their soft masses result in about $500~{\rm GeV}$. 
Alternatively, one can introduce a new gauge symmetry such as SU(2), which is not related to a mediation mechanism.
Then, $\{\hat{L},\hat{L}^{c};\hat{N},\hat{N}^{c}\}$ are regarded as the SU(2) doublets.
Here, we comment on the latter case.

The superpotential of the Higgs sector with such fields is given by
\dis{ \label{Higgs}
W_{\rm Higgs}=
y_h\hat{L} h_u\hat{N}^c +y_h^\prime \hat{L}^ch_d\hat{N}+\hat{\mu}_L\hat{L} \hat{L}^{c}+\hat{\mu}_N\hat{N} \hat{N}^{c}+\hat{\mu}_H\hat{N}_H \hat{N}_H^{c} ,
}
where $y_h^{(\prime)}$ and $\hat{\mu}_{L,N,H}$ ($\hat{\mu}_L\gtrsim\hat{\mu}_N$) are dimensionless and dimensionful parameters.
$\{\hat{N}_H,\hat{N}_H^{c}\}$ are also SU(2) doublets, but neutral under the SM gauge symmetry.
They are the spontaneous breaking sector of SU(2): their negative soft mass squareds
are assumed, breaking SU(2) above TeV scale.
For the gauge coupling unification, two pairs of $\{D^\prime,D^{c\prime}\}$ need to be accompanied
with their relatively heavy mass terms ($\gtrsim$ a few TeV), even if we do not write them down explicitly in \eq{Higgs}.
They are neutral under SU(2).
Thus, we have three pairs of $\{{\bf 5},\overline{\bf 5}\}$ in total (i.e. $v=3$), including one pair of $\{{\bf 5},\overline{\bf 5}\}$ introduced in Table II for the mixing of the SM chiral fermions.

Since $\{\hat{L},\hat{N}^c\}$ couple to the Higgs $h_u$, they make contributions to the radiative Higgs mass ($\equiv\Delta m_h^2$) as well as the renormalization of the soft mass squared of $h_u$ ($\equiv\Delta m_2^2$) \cite{KS2}:
\dis{
&~~ \Delta m_h^2|_{\rm new}\approx n_c\frac{|y_h|^4}{4\pi^2}v_h^2{\rm sin}^4\beta
~{\rm log}\left(\frac{{M}^2+\widetilde{m}_l^2}{{M}^2}\right) ,
\\
&\Delta m_2^2|_{\rm new}\approx n_c\frac{|y_h|^2}{8\pi^2}
\bigg[f_Q({M}^2+\widetilde{m}_l^2)-f_Q({M}^2)\bigg]_{Q=M_G} ,
}
where $n_c=2$ for the SU(2) doublets, $v_h$ ($\approx 174~{\rm GeV}$) indicates the Higgs VEV,  and $f_Q(m^2)$ is defined as $f_Q(m^2)\equiv m^2\{{\rm log}(\frac{m^2}{Q^2})-1\}$. 
$\widetilde m_l^2$ denotes the soft mass squared of the extra vector-like leptons, and $M^2$ is the mass squared of a fermionic component, $M^2\approx |\hat\mu_L|^2+|y_h|^2v_h^2\sin^2\beta$.
Note that $\Delta m_h^2|_{\rm new}$ is proportional to $|y_h|^4\times v_h^2\sin^4\beta$, 
while $\Delta m_2^2|_{\rm new}$
is to $|y_h|^2\times\{M^2+\widetilde{m}_l^2,M^2\}$.
Similarly, one could consider the radiative Higgs mass by $\{\hat{L}^c,\hat{N}\}$ via the $y_h^\prime$ coupling in \eq{Higgs}. However, its contribution would be proportional to $\cos^4\beta$, and so it 
is relatively suppressed for $\tan\beta\gtrsim 1$.  
$\Delta m_2^2|_{\rm new}$ is eventually associated with the fine-tuning issue, because it affects  determination of the $Z$ boson mass. 
The {\it quartic} power of $y_h$ in $\Delta m_h^2|_{\rm new}$ makes the radiative Higgs mass very efficiently raised. 
In order to raise the radiative Higgs mass, but holding the fine-tuning, hence, a larger $y_h$ and smaller masses $\{\widetilde{m}_l^2,|\hat{\mu}_L|^2\}$ need to be taken. 
Since the experimental bound is not severe yet, relatively light vector-like leptons are still conceivable. 

As shown in Ref.~\cite{KS2}, in the case of three pairs of $\{{\bf 5},\overline{\bf 5}\}$,
the {\it maximally allowed} $y_h$ at the EW scale can reach 1.78 for ${\rm tan}\beta\gtrsim 2$,
which makes 126 GeV Higgs mass easily explained, avoiding the Landau-pole problem.
%
%
For the masses of the new vector-like leptons heavier than 470 GeV (440 Gev) with ${\rm tan}\beta = 10$ ($50$), 
the oblique parameters $(\Delta S,\Delta T)$ induced by the new vector-like leptons become inside the 1$\sigma$ band. 
Only if $\hat{\mu}_L\gtrsim\hat{\mu}_N$, the charged components of $\{\hat{L},\hat{L}^c\}$ produced in the collider can immediately decay into $\{\hat{N},\hat{N}^c\}$, the neutralino, and SM fermions. 

%
%
%
%
%
%
%
%

\section{Conclusion}

We have constructed the U(1)$_{B_1+B_2-2L_1}$ mediated SUSY breaking model,
in which the first two generations of squarks ($\approx 7~{\rm TeV}$) and the first generation of sleptons ($\approx 40~{\rm TeV}$) can be made quite heavier than the other SUSY particles.
Hence, non-observation of SUSY particles at the LHC
and FCNC associated with the electron such as $\mu^-\rightarrow e^-\gamma$ can  easily be understood in this framework.
The discrepancy of $(g-2)_\mu$ can be explained
with the relatively light smuon/sneutrino and chargino/neutralino ($\approx 400~{\rm GeV}$).
The fine-tuning in the Higgs sector associated with the stop can be relieved, since the stop mass is relatively light ($\gtrsim 500~{\rm GeV}$) in this model.
Two-loop effects by the heavy sfermions can protect the small stop and smuon/sneutrino masses
against the quantum correction by the heavy gluino ($\approx 1.2~{\rm TeV}$).
By introducing extra vector-like matter, the radiative corrections to the Higgs mass can be enhanced up to 126 GeV,
and the desired mixings among the SM chiral fermions
can be generated after U(1)$_{B_1+B_2-2L_1}$ breaking.

\acknowledgments

We thank Jihn E. Kim for valuable discussions and comments. 
J.-H.H. is supported in part by DOE grant DE-FG02-13ER42022.
B.K. is supported by 
the National Research Foundation of Korea (NRF) funded by the Ministry of Education, Grant No. 2013R1A1A2006904, and also in part 
by Korea Institute for Advanced Study (KIAS) grant funded by the Korean government.
This study was supported by the Research Fund Program of Research Institute for Basic Sciences, 
Pusan National University, Korea, 2009, Project No. 
RIBS-PNU-2010-303. 



\end{document}